\pdfoutput=1
\documentclass[prd,nofootinbib,twocolumn,superscriptaddress,longbibliography]{revtex4-1}
\usepackage{booktabs}
\usepackage{multirow}
\usepackage{hyperref} 
\usepackage{graphicx}
\usepackage{dcolumn}
\usepackage{bm}
\usepackage{amsmath}
\usepackage{mathtools}
\usepackage{amsfonts}
\usepackage{pbox}
\usepackage{color}
\usepackage{xcolor}
\usepackage{tabularx}
\usepackage{subfigure}
\usepackage{graphicx}
\usepackage{array}
\usepackage{psfrag}
\usepackage{lipsum}
\usepackage{enumitem}  
\usepackage{changes}
\usepackage{soul}
\usepackage{float}
\usepackage{hhline}
\usepackage{xcolor}
\usepackage{bm,color,psfrag}
\usepackage{lipsum}
\usepackage[skip=1ex]{caption}
\usepackage{enumitem}  
\usepackage{amsfonts,amssymb}

\definecolor{MyDarkBlue}{rgb}{0,0.1,0.7}

\hypersetup{pdfborder={0 0 0},colorlinks,breaklinks=true,
  urlcolor={MyDarkBlue},citecolor={MyDarkBlue},linkcolor={MyDarkBlue}}

\def\equationautorefname~#1\null{Eq.(#1)\null}

\begin{document}

\title{Dynamical analysis of logarithmic energy-momentum squared gravity }

\author{Giovanni Acquaviva}
\email{gioacqua@gmail.com}
\affiliation{Arquimea Research Center, Camino de las Mantecas, 38320, Santa Cruz de Tenerife, Spain}
\author{Nihan Kat{\i}rc{\i}}
\email{nkatirci@dogus.edu.tr}
\affiliation{Department of Electrical and Electronics Engineering Do\u gu\c s University, \" Umraniye, 34775 Istanbul, Turkey}

\begin{abstract}
We perform the dynamical system analysis of a cosmological model in the energy-momentum squared gravity (EMSG) of the form $f(T_{\mu\nu} T^{\mu\nu})=\alpha\ln({\lambda}T_{\mu\nu} T^{\mu\nu})$, which is known as energy-momentum log gravity (EMLG). In particular, we show that the analytical cosmological solution of EMLG presented by Akarsu {\it et al.} (Eur. Phys. J. C 79:846, 2019) is a future attractor. It includes new terms in the right-hand side of the Einstein field equations, which yield constant inertial mass density and provide a dynamical dark energy with a density passing below zero at large redshifts, accommodating a mechanism for screening $\Lambda$ in the past for $\alpha<0$, suggested for alleviating some cosmological tensions. We show
that the second law of thermodynamics requires $\alpha\leq0$ that allows the screening mechanism to take place. We also show that the model gives rise to an entire class of new stable late-time solutions with  $H\rightarrow\sqrt{(\Lambda+2\alpha)/3}$ as $a\rightarrow\infty$, where the new term is due to the constant effective inertial mass density that arises from EMLG contribution of dust, whereas $H\rightarrow\sqrt{\Lambda/3}$ as $a\rightarrow\infty$ in the $\Lambda$CDM model. We also show the existence of new interesting features and trajectories that are absent  in $\Lambda$CDM with or without spatial curvature.
\end{abstract}
\date{\today}

\maketitle
\section{Introduction}
In general relativity (GR), the momentum conservation equation is given by ${\rm D}^\mu p+\varrho\dot{u}^\mu=0$, where $\rho$ and $p$ are correspondingly energy density and pressure, $\dot{u}^\mu$ is the four-acceleration, and ${\rm D}_\nu$ is the spatial gradient (the covariant derivative operator orthogonal to $u^\mu$) defined as ${\rm D}_\nu f= \nabla_\nu f+u_\mu\dot f$, a dot denotes derivative with respect to the proper time $t$. Analogously to Newton's second law of motion, $\varrho=\rho+p$ is the \textit{inertial mass density}, as it is the multiplier of the four acceleration \cite{EllisRC,Ellis:1998ct}. The inertial mass density of perfect fluids described by an equation of state (EoS) parameter $w=p/\rho$ is then expressed by $\varrho=(1+w)\rho$. Note that inertial mass density and energy density are identically the same, $\varrho=\rho$, for a non-relativistic source, viz., dust, as expected. On the other hand, it is simply null for the usual vacuum energy of Quantum Field Theory (described by the EoS of the form $p=-\rho$), which may constitute a better way of describing the vacuum energy as opposed to the energy density, which is a constant subject to observations.  The inertial mass density of a scalar field is then described by $\varrho_{\phi}=\rho_{\phi}+p_{\phi}=\dot{\phi}^2$: hence the constant (or very slowly rolling) scalar field gives rise to a null inertial mass density like the one that of the usual vacuum energy (or the cosmological constant, $p=-\rho$). Phenomenological generalizations of the usual vacuum energy based on the fact that it yields zero inertial mass density, $\varrho=0$, have been studied in recent literature to address some cosmological tensions; see, for instance, Ref.~\cite{Acquaviva:2021jov} (see also \cite{Bouhmadi-Lopez:2014cca}), which considers  a minimal deviation from the zero inertial mass density by promoting it to an arbitrary constant ($\varrho=\rm const$), called \textit{simple graduated dark energy} (simple-gDE)\footnote{In this study we consider possible isotropic extensions when the anisotropy in expansion is allowed, it is also possible to introduce \textit{deformed vacuum energy} preserving zero inertial mass density via its anisotropic pressure, see \cite{Akarsu:2020pka}. Since it will reveal more on the nature of DE,  anisotropic expansion searches occupy an important place in the upcoming missions such as the Euclid satellite~\cite{Amendola:2016saw}.
}; or Ref.~\cite{Akarsu:2019hmw} (see also Refs.~\cite{Barrow:1990vx,Stefancic:2004kb,Akarsu:2021fol,Ozulker:2022slu}), which considers the minimal dynamical deviation from the null inertial mass density (corresponding to the minimal deviation from simple-graduated dark energy) of the form $\varrho \propto\rho^{\lambda}<0$ (where $\lambda$ is a ratio of two odd integers). 

In the simple-gDE, negative values of the constant inertial mass density (which implies a violation of the null energy condition, $\varrho\geq0$) result in an energy density that decreases with increasing redshift, like in the phantom dark energy models, alas with a difference, as it crosses below zero at a certain redshift. This model has been constrained in Ref.~\cite{Acquaviva:2021jov} by the joint observational data [Planck cosmic microwave background (CMB), baryon acoustic oscillations (BAO), Type Ia Supernovae (SNIa), and cosmic chronometers (CC)] and found that the data favor its positive values, namely, $\varrho= (3.06\pm2.28)\times 10^{-31} \, {\rm g\, cm^{-3}}$ (at the order of $\mathcal{O}(10^{-12})\,\rm eV^4$) at 68\% confidence level, rather than zero, which would correspond to the usual vacuum energy. Although, with this value, the dynamics of the universe does deviate only insignificantly from the $\Lambda$CDM model in the past, the dynamical analysis carried out in the same paper has shown that for $\varrho>0$ the dynamics leads to a recollapsing future; only for $\varrho<0$ the asymptotic de Sitter future is an attractor like in the standard $\Lambda$CDM model.

Generation of new terms in the Einstein field equations that yield a constant inertial mass density is possible in Lagrangian-based theories using usual cosmological fluids as well, rather than a phenomenological generalization of the usual vacuum energy. For instance, such terms that yield constant inertial mass density arise from barotropic perfect fluid via the energy-momentum squared gravity (EMSG) of the form $f(T_{\mu\nu}T^{\mu\nu})\propto \ln(\lambda\,T_{\mu\nu}T^{\mu\nu})$, known as energy-momentum log gravity (EMLG), see Ref.~\cite{Akarsu:2019ygx} (see also \cite{Board:2021gnj}). There is a rapidly growing  literature considering EMSG, see \cite{Roshan:2016mbt,Akarsu:2017ohj,Board:2017ign,Akarsu:2018zxl,Akarsu:2018aro,Akarsu:2020vii,Yousaf:2021xex,Barbar:2019rfn}
for some cosmological applications and \cite{Faria:2019ejh,Chen:2019dip,Nazari:2020gnu,Rudra:2020rhs,Chen:2021cts} for some other applications.
   EMSG theories are a special family of a Lorentz invariant and covariant generalization of GR proposed in Ref.~\cite{Katirci:2014sti} obtained by adding a term $f(T_{\mu\nu}T^{\mu\nu})$ to the Einstein-Hilbert (EH) action of GR.  Such generalizations of GR include a new type of contributions of the material stress to the right-hand side of the Einstein field equations, without invoking some new type of sources (for other similar type of theories, see, e.g. \cite{Harko:2010mv,Harko:2011kv}). In the framework of EMSG theories, EMLG is the one and only choice that generates an additional constant energy density term that accompanies the matter energy density ($\rho$) without contributing to the pressure equation.\footnote{Due to the presence of the logarithmic function of the trace of EMT, similar generation of constant inertial mass density is also possible in $f(R,T)$ gravity of the form 
   $f(R,T)=R+2\beta\ln{T}$, proposed in \cite{elizalde}, yet it has not been mentioned by the authors [see Eqs.(11) and (12)].}
   
   The dynamical analysis of $f(R,\bf{T^2})$ theories has been studied in \cite{Bahamonde:2019urw}. The authors consider $f(R,{\bf T^{2}})=\beta R^{n}+\zeta({\bf T^{2}})^{m}$ as a minimally coupled form of $f(R,\bf{T^2})$, but their analysis does not contain the case $n=1$, the simplest extension of GR --- staying loyal to EH action --- via a non-linear matter modification, viz., $f(R,{\bf T^{2}})=\beta R+\zeta({\bf T^{2}})^{m}$. Therefore, a dynamical analysis of EMSG theories --- in which the action takes the form $R+f({\bf T^{2}})$, i.e., the EH action plus a generic function of the self contraction of EMT both in powered form and in logarithmic form --- is still not available in the literature so far.
   
In this paper, we apply the methods of the dynamical system analysis to EMLG model~\cite{Akarsu:2019ygx} in the presence of a single perfect fluid. We discuss the constant (negative or positive) inertial mass density that arises from the usual cosmological fluids in EMLG by considering the stability of the critical points and some new possible trajectory consistent with theoretical expectations and observations, for spatially flat and curved spacetimes.

The strength of the EMLG modification on top of the EH is controlled by a constant $\bar{\alpha}$. It has been found in~\cite{Akarsu:2019ygx} that the data favor slightly negative values of $\bar{\alpha}$ (yet not inconsistent with zero, which is the GR limit of the model), viz., $\bar{\alpha}=-0.032\pm0.043$, implying the possibility of a negative valued constant effective inertial mass density which arises in the presence of dust, $\varrho_{\rm m, emlg}=\bar{\alpha}\rho_{\rm m0}$ with $\rho_{\rm m0}>0$. Accordingly, the new contributions of dust to the modified Friedmann equations mimic a source with a constant inertial mass density and provide a dynamical dark energy with a density passing below zero at large redshifts. That is, this model accommodates a mechanism for screening $\Lambda$ in the past, suggested for alleviating some cosmological tensions. It has been reported that the $H_0$ tension --- as well as a number of other low-redshift discrepancies --- may be alleviated by a dynamical dark energy that assumes negative or vanishing energy density values at high redshifts \cite{Delubac:2014aqe,Sahni:2014ooa,Aubourg:2014yra,Capozziello:2018jya,Wang:2018fng,Poulin:2018zxs,Dutta:2018vmq,Banihashemi:2018oxo,Visinelli:2019qqu,Akarsu:2019hmw,DiValentino:2020naf,Akarsu:2021fol,Escamilla:2021uoj}. See also Ref. \cite{DiValentino:2020zio} for a list of references on the $H_0$ tension and Ref. \cite{DiValentino:2021izs} for a review on its possible solutions. See \cite{Abdalla:2022yfr} for a review on cosmological tensions and possible solutions.

The constraints on the Hubble constant along with their errors at the $68\%$ and $95\%$ confidence levels for the EMLG and the $\Lambda$CDM models are $H_0=68.20\pm2.13\pm4.15\, {\rm km s^{-1} Mpc^{-1}}$ and $H_0=66.86\pm0.90\pm1.74\, {\rm km s^{-1} Mpc^{-1}}$, respectively (see \cite{Akarsu:2019ygx} for a detailed discussion). In this study, we discuss the signature of the inertial mass density associated with the source, analyzing the entire global dynamics of these models, thereby complementing the observational analysis results in the literature.

\section{Energy-Momentum Log Gravity}
\label{sec:EMLG}
The modification of the form $f\left(T_{\mu\nu}T^{\mu\nu}\right)$ in the EH action with a cosmological constant $\Lambda$ is written as \cite{Katirci:2014sti}
\begin{equation}
\begin{aligned}
S=&\int {\rm d}^4x \sqrt{-g}\,\left[\frac{1}{2\kappa}\left(R-2\Lambda \right)+f\left(T_{\mu\nu}T^{\mu\nu}\right)+\mathcal{L}_{\rm m}\right],
\label{action}
\end{aligned}
\end{equation}
where $R$ is the scalar curvature, $g$ is the determinant of the metric, and $\mathcal{L}_{\rm m}$ is the Lagrangian density corresponding to the matter source that will be described by the EMT, 
  \begin{align}
  \label{tmunudef}
 T_{\mu\nu}=-\frac{2}{\sqrt{-g}}\frac{\delta(\sqrt{-g}\mathcal{L}_{\rm m})}{\delta g^{\mu\nu}}=g_{\mu\nu}\mathcal{L}_{\rm m}-2\frac{\partial \mathcal{L}_{\rm m}}{\partial g^{\mu\nu}},
 \end{align}
which depends only on the metric tensor components, and not on its derivatives.
Here, the cosmological constant $\Lambda$ is considered as a bare cosmological constant in accordance with the Lovelock's  theorem \cite{Lovelock:1971yv,Lovelock:1972vz}, stating that $\Lambda$ arises as a constant of nature like Newton's gravitational constant $G=\kappa/8\pi$. 

The field equations for a generic function of the self contraction of EMT read as follows:
  \begin{equation}
  \label{eq:fieldeq1}
G_{\mu\nu}+\Lambda g_{\mu\nu}= \kappa\left(T_{\mu\nu}+f g_{\mu\nu}-2\frac{\partial{f}}{\partial (T_{\mu\nu}T^{\mu\nu})}\theta_{\mu\nu}\right),
\end{equation} 
where $G_{\mu\nu}=R_{\mu\nu}-\frac{1}{2}Rg_{\mu\nu}$ is the Einstein tensor, and the new tensor $\theta_{\mu\nu}$ is defined as 

\begin{align}
\theta_{\mu\nu}=&T^{\sigma\epsilon}\frac{\delta T_{\sigma\epsilon}}{\delta g^{\mu\nu}}+T_{\sigma\epsilon}\frac{\delta T^{\sigma\epsilon}}{\delta g^{\mu\nu}}\nonumber\\
=&-2\mathcal{L}_{\rm m}\left(T_{\mu\nu}-\frac{1}{2}g_{\mu\nu}T\right)-T T_{\mu\nu}\nonumber\\
&+2T_{\mu}^{\gamma}T_{\nu\gamma}-4T^{\sigma\epsilon}\frac{\partial^2 \mathcal{L}_{\rm m}}{\partial g^{\mu\nu} \partial g^{\sigma\epsilon}},
\label{theta}
\end{align}
where $T$ is the trace of the EMT. One may immediately notice that, irrespective of the functional choice for $f$, the term $\kappa fg_{\mu\nu}$ in Eq.~\eqref{eq:fieldeq1} resembles the cosmological constant term $\Lambda g_{\mu\nu}$, yet it will turn out that it is dynamical in character.

We consider an isotropic and homogeneous spacetime described by the Robertson-Walker metric:
\begin{equation}
\label{RW}
{\rm d}s^2=-{\rm d}t^2+a^2\,\left[\frac{{\rm d}r^2}{1-kr^2}+r^2({\rm d}\theta^2+\sin^2\theta {\rm d}\phi^2)\right],
\end{equation}  
where the scale factor $a=a(t)$ is a function of cosmic time and the spatial curvature parameter $k$ takes values in $\{-1,\,0,\,1\}$ corresponding to open, flat and closed 3-spaces respectively. The matter source is described by a perfect fluid with Lagrangian $\mathcal{L}_{\rm m}=p$, such that the EMT appearing in Einstein field equations as
\begin{align}
\label{eq:em}
T_{\mu\nu}=(\rho+p)u_{\mu}u_{\nu}+p g_{\mu\nu},
\end{align} 
giving  
\begin{align}
\label{eq:trace}
T_{\mu\nu}T^{\mu\nu}&=\rho^2(3w^2+1),\\
\theta_{\mu\nu}&=-\rho^2(3w+1)(w+1)u_{\mu}u_{\nu}.
\label{eq:thetafrw}
\end{align}
As the definition of the matter Lagrangian that gives rise to the perfect fluid EMT is not unique, one could choose either $\mathcal{L}_{\rm m}=p$ or $\mathcal{L}_{\rm m}=-\rho$, which would result in the same EMT, viz., the $T^{\mu\nu}$ that describes perfect fluid matter distributions as given in \eqref{eq:em}, see \cite{Bertolami:2008ab,Faraoni:2009rk} for a detailed discussion. In the present study, following the literature to date on EMSG theories, we consider $\mathcal{L}_{\rm m}=p$; in this case, the last term in \eqref{theta} vanishes.\footnote{A choice of $\mathcal{L}_{\rm m}=-\rho$ would lead to differences at the level of equations of motion in EMSG; this point deserves further investigation.}

We see that the EMLG \cite{Akarsu:2019ygx}, viz., EMSG of the form
\begin{eqnarray}
\label{eq:fchoice}
f(T_{\mu\nu} T^{\mu\nu})=\alpha\ln({\lambda}T_{\mu\nu} T^{\mu\nu}),
\end{eqnarray}
is the one and only choice that leads to $-2\theta_{00}\partial{f}/\partial \mathbf{T}^2=2\alpha(3w^2+4w+1)/(3w^2+1)$ from the last term of Eq.~\eqref{eq:fieldeq1} for $\mu=\nu=0$, which implies an additional constant energy density term that accompanies the matter energy density ($\rho$) without contributing to the pressure equation. Here, $\lambda$ has the dimensions of inverse energy density squared so that the argument of the logarithm is dimensionless. The inertial mass density of the matter source ($\rho+p$) can be constant or dynamical, but the one arising from EMLG contributions, denoted with  the subscript ${\rm emlg}$, comes as a constant:
\begin{eqnarray}
\label{eq:imd0}
\varrho_{\rm emlg}=\rho_{\rm emlg}+p_{\rm emlg }=2\alpha\frac{(3w+1)(w+1)}{3w^2+1}.
\end{eqnarray}
 We note that this is a constant that depends on the EoS of the matter source; $\varrho_{\rm m,emlg}=2\alpha$ for dust ($w=0$), $\varrho_{\rm r,emlg}=4\alpha$ for radiation ($w=1/3$). It is worth observing that, as long as $\alpha\neq0$, the effective inertial mass density is null for the cases $w=-1$ and $-\frac{1}{3}$, corresponding to the conventional vacuum energy and cosmic strings. For positive (negative) values of $\alpha$, $\varrho_{\rm emlg}/2\alpha$ is positive (negative) for $w<-1$ and $w>-1/3$, and otherwise negative (positive).

From Eq.~\eqref{eq:fieldeq1}, we obtain the following pair of linearly independent modified Friedmann equations, for a single fluid cosmology,
\begin{align}
&3H^{2}+\frac{3k}{a^2}=\rho+\Lambda+\bar{\alpha} \rho_{0}+ \bar{\alpha} \rho_{0} \frac{2}{\gamma} \ln\left(\rho/\rho_{0}\right) , \label{eq:rhoprime}\\
&-2\dot{H}-3H^{2}-\frac{k}{a^2}= w \rho-\Lambda\nonumber\\
&\quad\quad\quad\quad\quad\quad\quad - \bar{\alpha} \rho_{0} \frac{2}{\gamma}\ln\left[\sqrt{3w^{2}+1}\;\left(\rho/\rho_{0}\right)\right],\label{eq:presprime}
\end{align}
where $\bar{\alpha}$ is a dimensionless constant defined by
\begin{align}
\label{gamma}
\bar{\alpha} = -\frac{\alpha\, \gamma}{\rho_0}, 
\end{align}
and
\begin{align}
\label{eq:gammadef}
 \quad \gamma = \ln\left( 3w^2+1  \right) - 2 \frac{(3w+1)(w+1)}{(3w^2+1)}.
\end{align}
The covariant divergence of the EMT becomes
\begin{equation}
\label{nonconservedenergy}
\nabla^{\mu}T_{\mu\nu}=\alpha\left[ g_{\mu\nu}\nabla^{\mu}\ln(\lambda\, T_{\sigma\epsilon}T^{\sigma\epsilon})
-2\nabla^{\mu}\left(\frac{\theta_{\mu\nu}}{T_{\sigma\epsilon}T^{\sigma\epsilon}}\right)\right],
\end{equation}
so that the corresponding local energy-momentum conservation equation \eqref{nonconservedenergy} reads
\begin{equation}
\label{noncons}
\dot{\rho}+3H(1+w)\rho\left[\frac{\rho-\frac{2\bar{\alpha} \rho_{0}(3w+1)}{\gamma(3w^2+1)}}{\rho+\frac{2 \bar{\alpha} \rho_{0}}{\gamma}} \right]=0. 
\end{equation}
The terms with $\bar{\alpha}$ in Eqs.~\eqref{eq:rhoprime} and \eqref{eq:presprime} give the EMLG-based constant inertial mass density as
\begin{align}
\label{eq:imd}
\varrho_{\rm emlg}=\rho_{\rm emlg}+p_{\rm emlg}=&\bar{\alpha} \rho_{0}\left[1-\gamma^{-1}\ln\left(3w^{2}+1\right)\right].
\end{align}

\section{DYNAMICAL ANALYSIS}
\label{sec:dyn}
In order to analyse the global behaviour of the model in the framework of cosmological dynamical systems, we should start by defining dimensionless variables and derive an autonomous system of equations for their evolution: this is usually done by normalizing over the squared expansion $H^2$, but in a generic curved FRW spacetime we have to take into account the possibility of bouncing and recollapsing models for which at some point $H=0$ and the dimensionless variables become singular. To obviate such problem, one can normalize instead over some quantity which is non-vanishing throughout the evolution.  In order to do so we define
\begin{equation}\label{eq:norm}
 D^2 \equiv H^2 + \frac{|k|}{a^2},
\end{equation}
and the new dimensionless variables
\begin{equation}
\begin{aligned}
\label{eq:var}
 \Omega = \frac{\rho}{3\, D^2}\quad &,\quad \Omega_0 = \frac{\rho_0}{3\, D^2}, \\
 \Omega_\Lambda = \frac{\Lambda}{3\, D^2}\quad &,\quad \Omega_H = \frac{H}{D}\, .
\end{aligned}
\end{equation}
The normalization $D$ has been introduced in \cite{Goliath:1998na} as a means to compactify the parameter space of spatially homogeneous cosmologies and include bouncing/recollapsing scenarios in the analysis, as is well defined throughout the whole cosmological evolution including possible turning points of the scale factor, see also \cite{Bahamonde:2018} for a comprehensive review of this and other methods to treat noncompact dynamical systems. We define as well a new evolution parameter related to the cosmic time $t$ by $d\tau = D\, dt$. It is possible now to obtain an autonomous system for such variables by taking the derivatives of the definitions above with respect to $\tau$, denoted here by a prime:
\begin{align}
  \label{eq:omegap}
 \Omega' &= -\, \Omega\, \Omega_H \Bigg[ 3\, (1+w)\, \frac{\gamma\, \Omega -2\, \bar{\alpha}\, \Omega_0\, \frac{3w+1}{3w^2+1}}{\gamma\, \Omega+2\, \bar{\alpha}\, \Omega_0}\Bigg. \nonumber \\
  &\quad\Bigg.+ 2\, \left( \frac{\dot{H}}{D^2} + \Omega_H^2 -1 \right) \Bigg],  \\
 \Omega_0' &= -2\, \Omega_0\, \Omega_H\, \left( \frac{\dot{H}}{D^2} + \Omega_H^2 -1 \right),  \\ 
 \Omega_H' &= \left( 1-\Omega_H^2 \right)\, \left( \frac{\dot{H}}{D^2} + \Omega_H^2 \right). 
 \label{eq:omegahp}
\end{align}

As is the case in standard cosmology as well as in some extensions, the Friedmann constraint defines a compact phase space, which means that all the critical points (CPs) of the system exist in a finite region. But in EMSG-type theories, one or more variables can change sign, see Eq.~\eqref{eq:rhoprime}, the phase space is not compact anymore and there probably will be CPs at infinity: the analysis of these requires a further redefinition of variable to compactify the space. Here, with this  \textit{clever} definition of variables, we are able to circumvent this issue from the beginning.

The evolution equation for $\Omega_\Lambda$ can be ignored thanks to the Friedmann constraint.  Note that in the limit $\bar{\alpha}\rightarrow0$ the dynamical equations for $\Omega$ and $\Omega_H$ decouple from $\Omega_0$ and constitute by themselves the usual cosmological dynamical system of GR.

In the system above we have not substituted $\dot{H}$ in terms of the dimensionless variables because the cases $k>0$ and $k\leq 0$ have to be treated separately, see e.g. \cite{Kerachian:2019tar}. The $o\Lambda$CDM model refers to the model including the spatial curvature ($k\neq0$) on top of the standard $\Lambda$CDM model.  In particular, depending on the sign of the curvature, we can recast Friedmann and Raychaudhuri equations respectively as follows: for positive spatial curvature (closed space) we have
  \begin{align}
    1 = \Omega + \Omega_\Lambda &+ \bar{\alpha}\, \Omega_0\, \left[ 1+\frac{2}{\gamma}\, \ln \left( \frac{\Omega}{\Omega_0} \right) \right], \label{eq:friedpos}\\
    \frac{\dot{H}}{D^2} + \Omega_H^2 =\ &1-\frac{3}{2}\, (1+w)\, \Omega\nonumber\\
    &- \frac{3}{2}\, \bar{\alpha}\, \Omega_0\, \left[ 1- \frac{2}{\gamma}\, \ln\sqrt{3w^2+1} \right], \label{eq:raypos}
  \end{align}
while for non-positive spatial curvature (open space) we have
  \begin{align}
    1 = 2\, \Omega_H^2\, -\, &\Omega - \Omega_\Lambda - \bar{\alpha}\, \Omega_0\, \left[ 1+\frac{2}{\gamma}\, \ln \left( \frac{\Omega}{\Omega_0} \right) \right], \label{eq:friedneg}\\
    \frac{\dot{H}}{D^2} + \Omega_H^2 =\ &2\, \Omega_H^2 - 1-\frac{3}{2}\, (1+w)\, \Omega \nonumber\\
    &- \frac{3}{2}\, \bar{\alpha}\, \Omega_0\, \left[ 1- \frac{2}{\gamma}\, \ln\sqrt{3w^2+1} \right]. \label{eq:rayneg}
  \end{align}
The $\dot{H}$ terms in the system Eqs.~\eqref{eq:omegap}-\eqref{eq:omegahp} can be substituted from Eq.~\eqref{eq:raypos} for the case $k>0$ or from Eq.~\eqref{eq:rayneg} for the case $k\leq0$.  Obviously, specializing in the spatially flat case $\Omega_H=\pm 1$, the equations for the two curvatures acquire the same form.

The Friedmann constraints, Eqs.~\eqref{eq:friedpos} and \eqref{eq:friedneg}, do not in general define compact parameter spaces because of the presence of the logarithmic terms, which can change sign depending on the relative magnitude of $\Omega$ and $\Omega_0$.

\begin{table*}[ht!]
\caption{Coordinates, cosmological parameters and stability features of the finite CPs of the system Eqs.~\eqref{eq:omegap}-\eqref{eq:omegahp}.}
\label{table:1a}
\centering 
\resizebox{14cm}{!}{
 \begin{tabular}{|c|c||c|c|c||c|c|c|c|}
  \hline
  $k$ & CP & $\Omega$ & $\Omega_0$  & $\Omega_{H}$ & $q$ & $w_{\rm eff}$ & Eigenvalues & Stability\\[0.1cm] 
 \hline\hline
 \multirow{6}{*}{\vspace{-7cm}$k=0$} & $A_0^+$ & 0 & 0 &  $1$ & $-1$ & $-1$ & $\left\{0 ,\ -2\ ,\ -3(1+w) \right\}$ & $\Bigg{\{}$\pbox[c][1.5cm]{2.5cm}{\text{sink} $w>-1$ \\ \text{saddle} $w<-1$} \\[0.1cm] 
 &$A_0^-$ & 0 & 0 &  $-1$ & $-1$ & $-1$ & $\left\{0 ,\ 2\ ,\ 3(1+w) \right\}$ & $\Bigg{\{}$\pbox[c][1.5cm]{2.5cm}{\text{source} $w>-1$ \\ \text{saddle} $w<-1$} \\[0.1cm] 
  & $B_0^+$ & $\frac{2\bar{\alpha}(1+3w)}{\gamma(1+3w^2)}\Omega_0$ & $\forall$ & $1$ & $-1$ & $-1$ & $\left\{0 ,\ -2\ ,\ -\frac{3(1+w)(1+3w)}{2+3w(1+w)} \right\}$ & $\Bigg{\{}$\pbox[c][2cm]{4cm}{\text{sink} $w<-1 , w>-1/3$ \\ \text{saddle} $-1<w<-1/3$} \\[0.1cm]
  & $B_0^-$ & $\frac{2\bar{\alpha}(1+3w)}{\gamma(1+3w^2)}\Omega_0$ & $\forall$ & $-1$ & $-1$ & $-1$ & $\left\{0 ,\ 2\ ,\ \frac{3(1+w)(1+3w)}{2+3w(1+w)} \right\}$ & $\Bigg{\{}$\pbox[c][2cm]{4cm}{\text{source} $w<-1 , w>-1/3$ \\ \text{saddle} $-1<w<-1/3$} \\[0.1cm]
    & $C_0^+$ & $1$ & 0 & $1$ & $(1+3w)/2$ & $w$ & $\left\{3(1+w) ,\ 3(1+w)\ ,\ (1+3w) \right\}$ & $\Bigg{\{}$\pbox[c][2.5cm]{4cm}{\text{source} $w>-1/3$ \\ \text{saddle} $-1<w<-1/3$ \\ \text{sink} $w<-1$} \\[0.1cm]
    & $C_0^-$ & $1$ & 0 & $-1$ & $(1+3w)/2$ & $w$ & $\left\{-3(1+w) ,\ -3(1+w)\ ,\ -(1+3w) \right\}$ & $\Bigg{\{}$\pbox[c][2.5cm]{4cm}{\text{sink} $w>-1/3$ \\ \text{saddle} $-1<w<-1/3$ \\ \text{source} $w<-1$} \\[0.1cm]
 \hline
 \multirow{4}{*}{\vspace{-3cm}$k<0$} & $A_-$ & $-\frac{2}{3(1+w)}+\frac{2 \bar{\alpha}(1+3w)}{\gamma(1+3w^2)}\Omega_0$ & $\forall$  & $0$ & $-2$ & $-5/3$ & $\textnormal{given along the text}$ & $\textnormal{given along the text}$ \\[0.1cm] 
 & $B_-$ & $-\frac{2}{3(1+w)}$ & $0$ & $0$ & $-2$ & $-5/3$ & $\left\{ 0,\ -i\sqrt{1+3w} ,\ i\sqrt{1+3w} \right\}$ & $\Bigg{\{}$\pbox[c][1.5cm]{2.5cm}{\text{saddle} $w<-1/3$ \\ \text{center} $w>-1/3$} \\[0.1cm]
 & $C_-^+$ & $0$ & $0$ & $\frac{1}{\sqrt{2}}$ & $0$ & $-1/3$ & $\left\{\sqrt{2} ,\ \sqrt{2} ,\ -\frac{1+3w}{\sqrt{2}} \right\}$ & $\Bigg{\{}$\pbox[c][1.5cm]{2.5cm}{\text{saddle} $w>-1/3$ \\ \text{source} $w<-1/3$} \\[0.1cm]
 & $C_-^-$ & $0$ & $0$ & $-\frac{1}{\sqrt{2}}$ & $0$ & $-1/3$ & $\left\{-\sqrt{2} ,\ -\sqrt{2} ,\ \frac{1+3w}{\sqrt{2}} \right\}$ & $\Bigg{\{}$\pbox[c][1.5cm]{2.5cm}{\text{saddle} $w>-1/3$ \\ \text{sink} $w<-1/3$} \\[0.2cm]
  \hline
  \multirow{2}{*}{\vspace{-1cm}$k>0$} & $A_+$ & $\frac{2}{3(1+w)}+\frac{2 \bar{\alpha}(1+3w)}{\gamma(1+3w^2)}\Omega_0$ & $\forall$  & $0$ & $0$ & $-1/3$ & $\textnormal{given along the text}$ & $\textnormal{given along the text}$ \\[0.1cm] 
  & $B_+$ & $\frac{2}{3(1+w)}$ & $0$ & $0$ & $0$ & $-1/3$ & $\left\{0 ,\ -\sqrt{1+3w}\ ,\ \sqrt{1+3w} \right\}$ & $\Bigg{\{}$\pbox[c][1.5cm]{2.5cm}{\text{saddle} $w>-1/3$ \\ \text{center} $w<-1/3$} \\[0.2cm]
\hline
\end{tabular}}
\end{table*}

\begin{figure*}[t!]
\captionsetup{justification=raggedright,singlelinecheck=false}
\par
\begin{center}
\subfigure[]{    
   \includegraphics[width=0.30\textwidth]{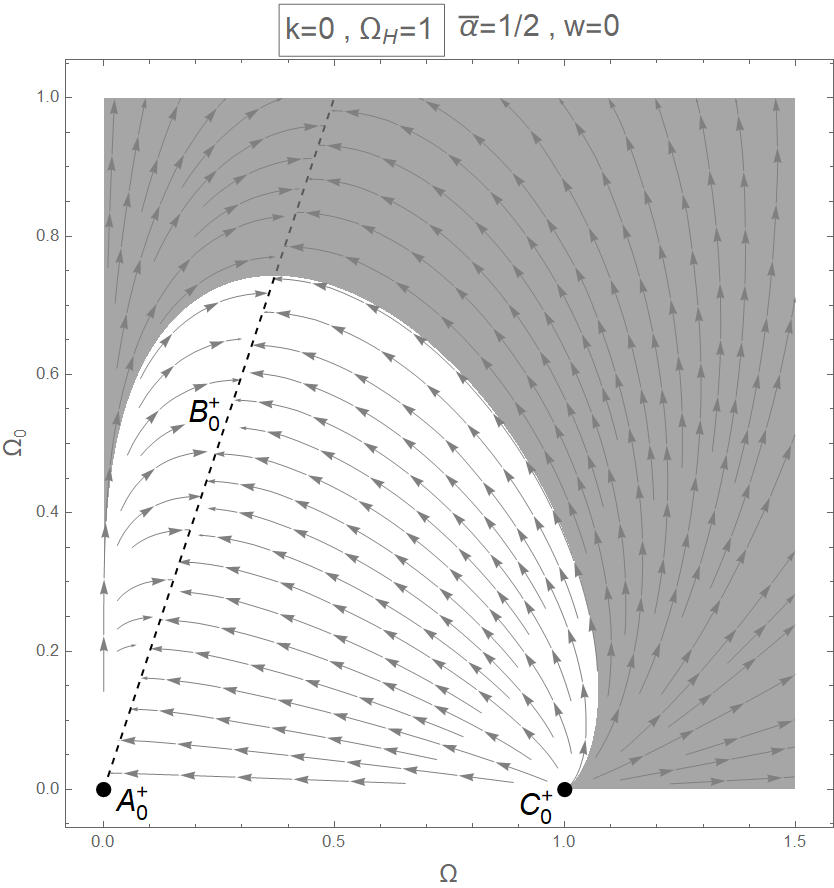}
   \label{fig:attpos}
            }
\subfigure[]{   
   \includegraphics[width=0.30\textwidth]{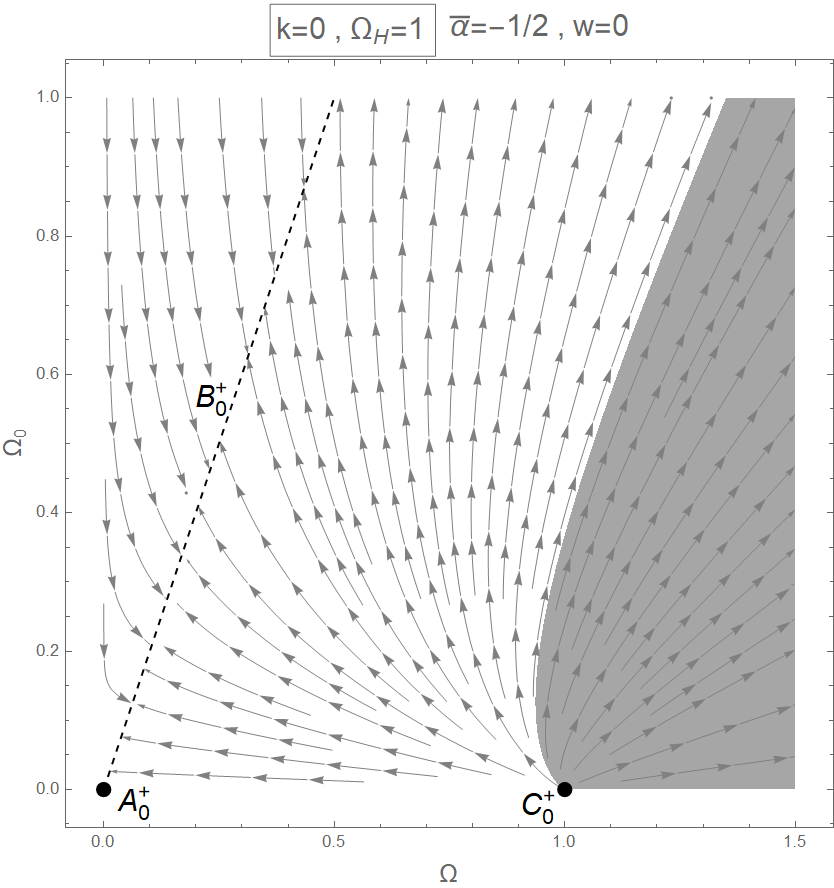}
   \label{fig:attneg}
           }
         
\end{center}
\caption{ Invariant subset corresponding to flat space in the expanding case ($\Omega_{H}=1$), with $w=0$ and (a)  $\bar{\alpha}=1/2$ and (b) $\bar{\alpha}=-1/2$. Critical elements are indicated with the black dots and the dashed line, while the shaded area is forbidden by the Friedmann constraint.  }
\label{fig:plotsflat}
\end{figure*}

The cosmological properties associated to the solutions listed above can be codified by the deceleration parameter $q$ and effective EoS parameter $w_{\rm eff}$ using the following expressions:
\begin{equation}
\label{eq:qdef}
 q = -1-\frac{\dot{H}}{H^2} = -1 -\Omega_H^{-2}\frac{\dot{H}}{D^2},
\end{equation}
and
\begin{equation}
\label{eq:weffdef}
 w_{\rm eff} =-1-\frac{2}{3}\frac{\dot{H}}{H^2} = -1-\frac{2}{3}\Omega_H^{-2}\frac{\dot{H}}{D^2},
\end{equation}
where the $\dot{H}/D^2$ in these expressions has to be substituted from Eqs.~\eqref{eq:raypos} and \eqref{eq:rayneg} according to the spatial curvature under consideration.  These quantities describe the effect of an effective fluid corresponding to the $T_{\mu\nu}$ contributions of EMLG.

\subsection{Critical Points}
\label{sec:fin_crit}
It is useful to identify first of all the invariant subsets of the system: these are $\Omega=0$, $\Omega_0=0$ and $\Omega_H=\pm 1$.  Initial conditions corresponding to such subspaces are left invariant by the evolution.  One can then uncover the presence of critical points by solving $\Omega'=\Omega_0'=\Omega_H'=0$, obtaining in this way up to twelve critical elements in the parameter space. Further, in order to assess the stability of the critical elements one has to calculate the Jacobian matrix evaluated in each point, obtain its eigenvalues $\lambda_i$ and inspect the sign of their real part: if Re$(\lambda_i)>0$ the associated eigendirection is unstable, whether Re$(\lambda_i)<0$ signals a stable one.  As a result, if all the eigenvalues have positive real part, the point is a source; if all the eigenvalues have negative real part, the point is a sink; mixed signs mean that the point is a saddle. Below, for the analysis of the cases with one vanishing eigenvalue we resort to a numerical inspection of the stability character. 

In what follows we present the mathematical and physical features of each one of the critical elements of this parameter space:

\begin{itemize}
    \item[$A_0^\pm$:] they describe exponentially expanding ($A_0^+$) or collapsing ($A_0^-$) spatially flat models, with $w_{\rm eff}=-1$ and $q=-1$, namely $\Lambda$-dominated solutions. For $w>-1$ the expanding solution is a sink while the contracting solution is a source; for $w<-1$ they are both saddle points.  These points are a special case of the critical line $B^{\pm}_0$ with $\Omega_0=0$ or $\bar{\alpha}=0$.
  
    \item[$B_0^\pm$:] exponentially expanding ($B_0^+$) or collapsing ($B_0^-$) solutions sourced by the modification $\bar{\alpha}$.  For $w<-1$ and $w>-1/3$ the expanding solution is a sink and the contracting solution is a source, while they are both saddle points otherwise. 
    
    The presence of $B_0^+$ as a future attractor means that different fluid sources, irrespective of $w$, contribute to the accelerated expansion with their inertial mass densities and approach de Sitter solution in the future (even in absence of $\Lambda$). In fact, the second term of the conservation equation \eqref{noncons} will be zero when we substitute the coordinate of this point, $\Omega=\frac{2\bar{\alpha}\Omega_{0}(3w+1)}{\gamma(3w^2+1)}$, leading a constant energy density, $\dot{\rho}=0$. For instance, for $w=0$, the energy density of dust does not reach zero asymptotically in the future, settling instead towards a non-vanishing constant: in fact, its inertial mass density $\Omega_{\rm m}\rightarrow -\bar{\alpha}\Omega_{\rm m0}\equiv\varrho_{\rm m}$. This solution of EMLG model\footnote{Here we analyze EMLG contributions without explicitly referring to dark energy; on the other hand, in \cite{Akarsu:2019ygx} the authors also consider all modifications (both new terms from EMLG and modified evolution of dust) as components of an effective dynamical ``dark energy", as it is properly defined in GR (see Sec. IV C of  \cite{Akarsu:2019ygx}).  It is possible to consider EMLG contributions as the only DE contributions: these appear to couple non-minimally with dust due to the modified redshift dependence, but actually dust and DE only couple to the spacetime governed by EMLG.  Therefore, in  \cite{Akarsu:2018aro}, in the context of scale-invariant EMSG theory, authors name it  as pseudo-interaction in dust and DE.  Moreover, interacting Dark Energy models have been used to try and resolve the H0 tension \cite{Kumar:2021eev}.} , shown in Figure \ref{fig:plotsflat}, has been presented in \cite{Akarsu:2019ygx} with a screening of dark energy via dust in EMLG. See Section \ref{sec:discussion} for a detailed discussion. 

    \item[$C_0^\pm$:] spatially flat, perfect fluid dominated solutions, with $w_{\rm eff}=w$  and $q=(1+3w)/2$. For $w>-1/3$ the expanding solution ($C_0^+$) is a past source while the contracting one ($C_0^-$) is a future sink; their behaviour is inverted for $w<-1$.
    
\begin{figure*}[t!]
\captionsetup{justification=raggedright,singlelinecheck=false}
\par
\begin{center}
\subfigure[]{    
   \includegraphics[width=0.21\textwidth]{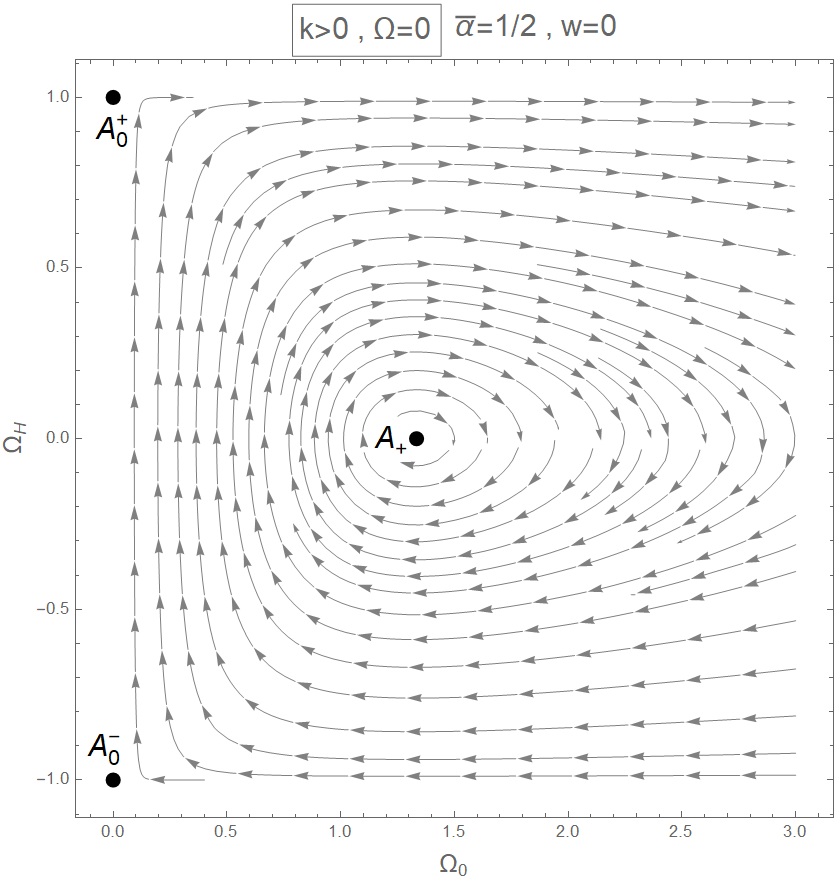}
            }
\subfigure[]{   
   \includegraphics[width=0.21\textwidth]{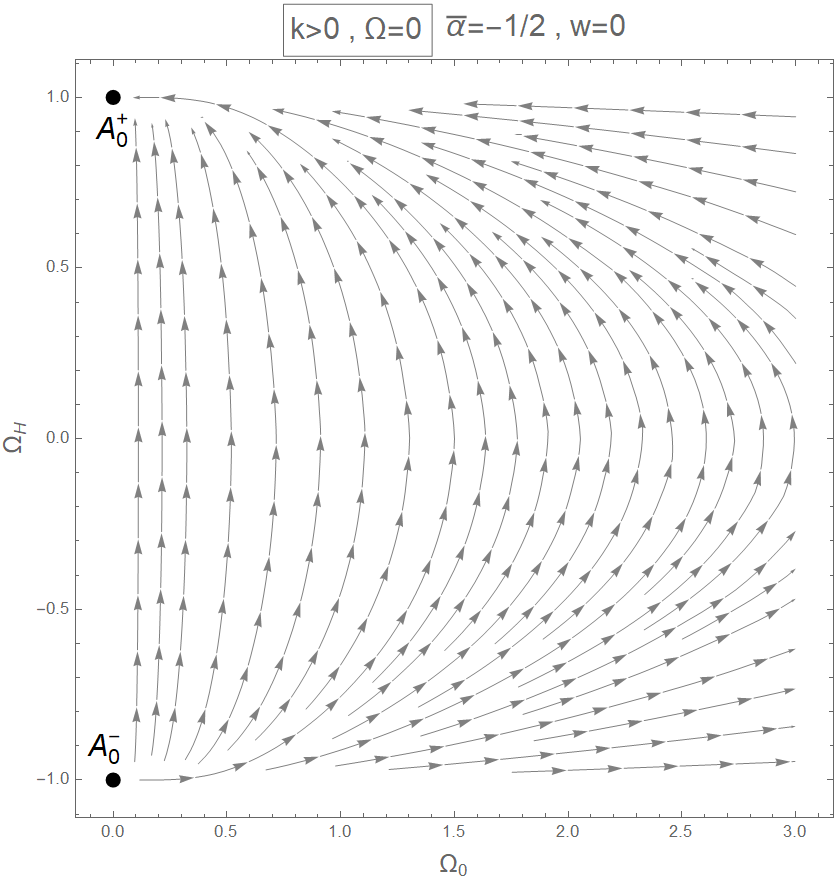}
           }
           \subfigure[]{   
   \includegraphics[width=0.21\textwidth]{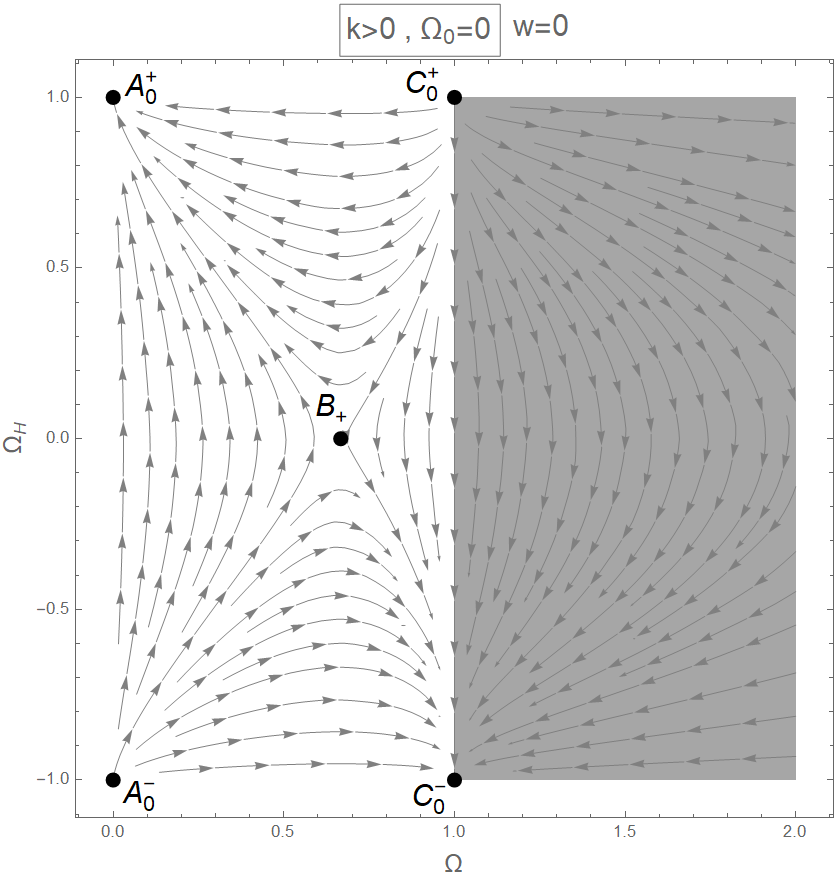}
           }
            \subfigure[]{ 
           \includegraphics[width=0.23\textwidth]{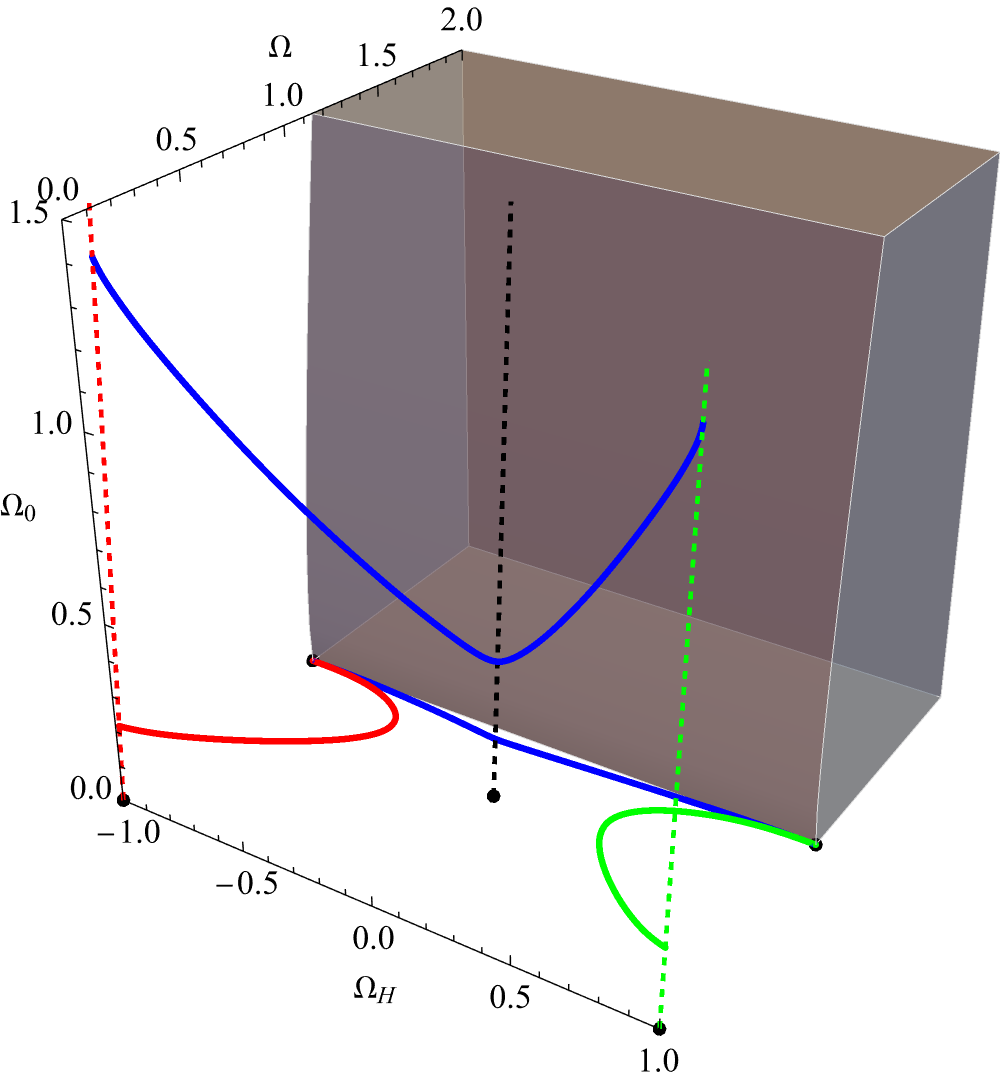}
           }
\end{center}
\caption{  Invariant subsets corresponding to positive curvature space for dust ($w=0$) with (a) $\Omega=0$ and $\bar{\alpha}=1/2$, (b) $\Omega=0$ and $\bar{\alpha}=-1/2$, and (c) $\Omega_0=0$. (d) A portion of the parameter space for the case of positive curvature for $\bar{\alpha}=-0.075$, corresponding to the lower observational bound on $\bar{\alpha}$ where the upper bound is  $\bar{\alpha}=0$ ($o\Lambda$CDM). The grey region is forbidden by the Friedmann constraint. We notice ever expanding (green), ever collapsing (red), and both bouncing and recollapsing trajectories (blue).}
\label{fig:plotspos}
\end{figure*}
\begin{figure*}
\captionsetup{justification=raggedright,singlelinecheck=false}
\par
\begin{center}

\subfigure[]{ 
   \includegraphics[width=0.21\textwidth]{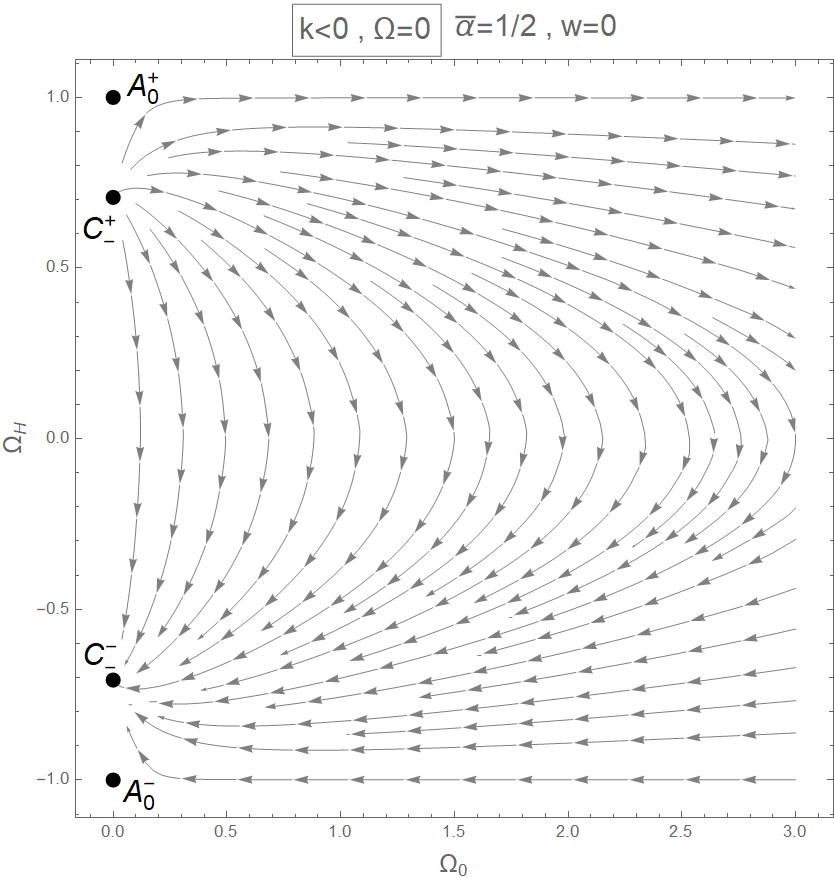}
            } 
\subfigure[]{   
   \includegraphics[width=0.21\textwidth]{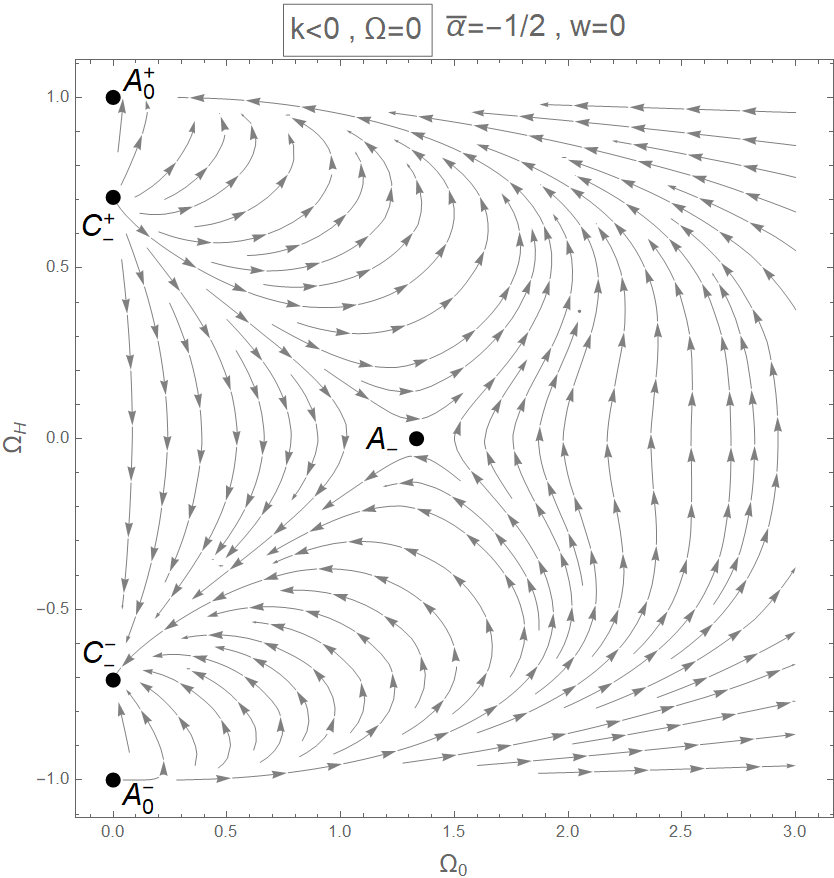}
           }
           \subfigure[]{   
   \includegraphics[width=0.21\textwidth]{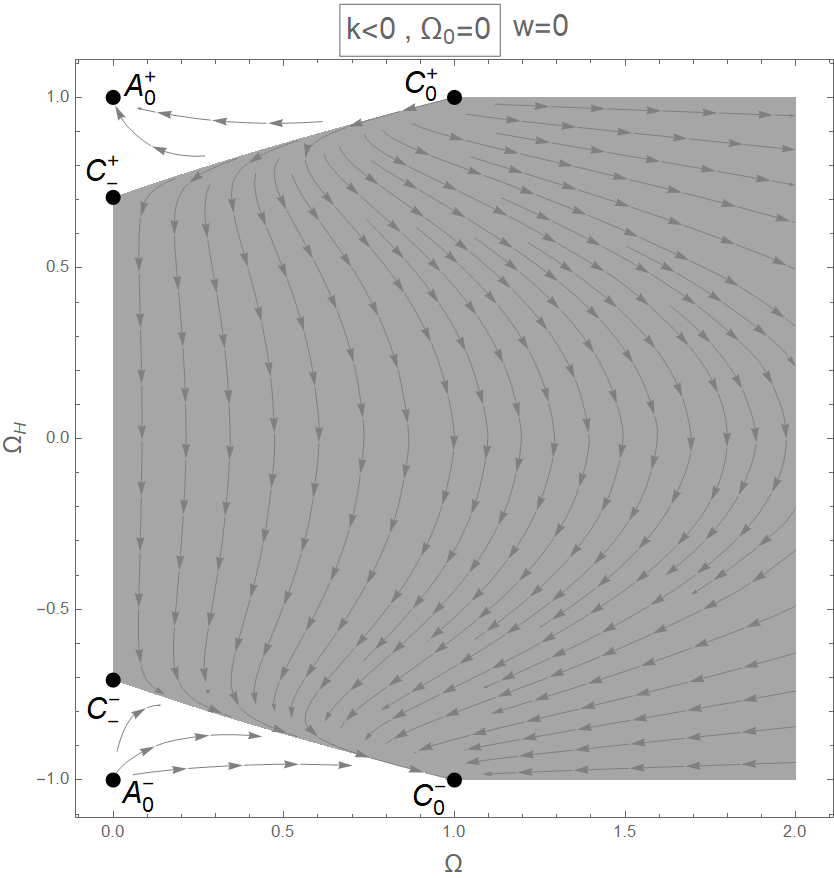}
    \label{fig:c}
           }
            \subfigure[]{ 
           \includegraphics[width=0.23\textwidth]{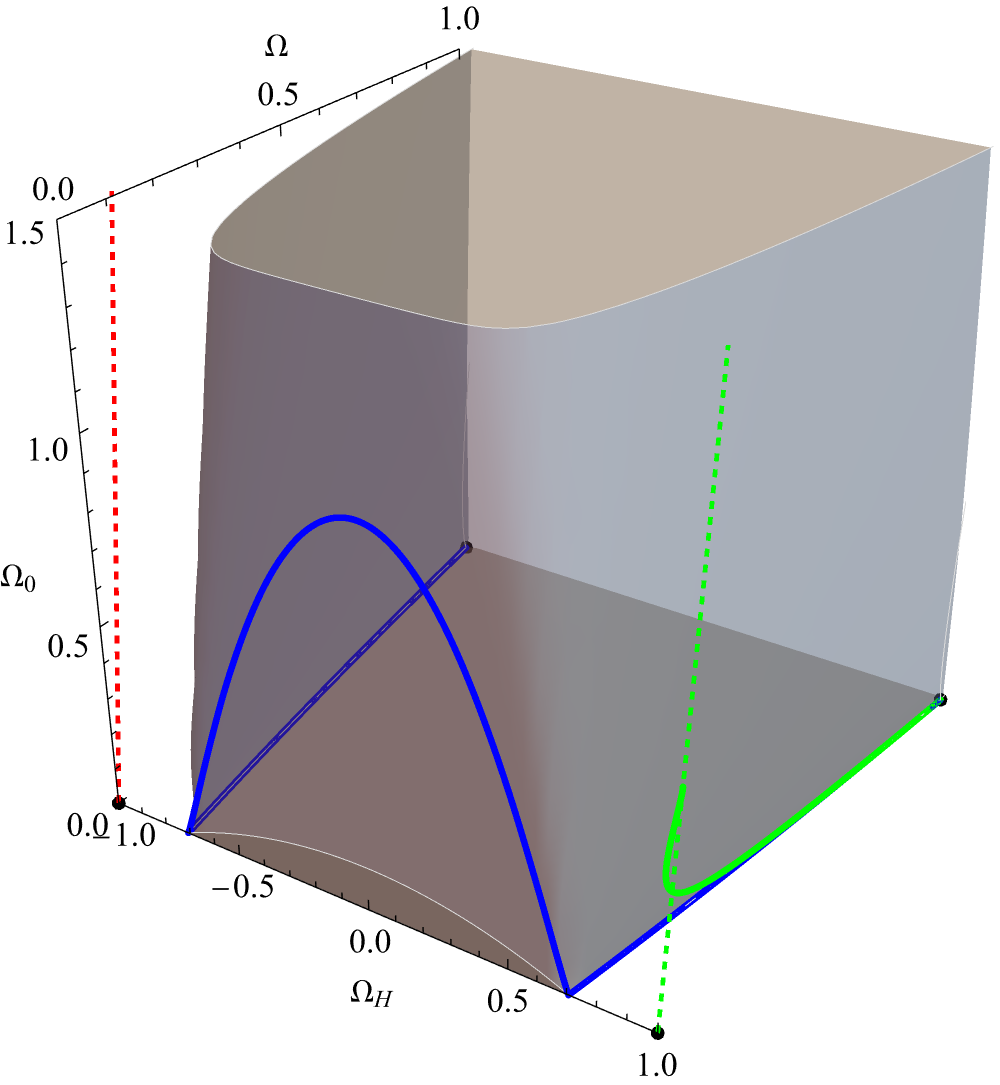}
           \label{fig:d}
             }
           
\end{center}
\caption{ Invariant subsets corresponding to negative curvature space for dust ($w=0$) with (a) $\Omega=0$ and $\bar{\alpha}=1/2$, (b) $\Omega=0$ and $\bar{\alpha}=-1/2$, (c) $\Omega_0=0$ and (d) A portion of the parameter space for the case of negative curvature with $\bar{\alpha}=-0.075$, corresponding to the lower observational bound on $\bar{\alpha}$ where the upper bound is  $\bar{\alpha}=0$ ($o\Lambda$CDM). Apart from the expanding solution that goes towards a de Sitter future (green), notice that, differently from GR, in this case there are trajectories corresponding to recollapsing models (blue). }
\label{fig:plotsneg}
\end{figure*}

    \item[$A_-$:] a critical line representing negative curvature-dominated solutions. If we specialize to the case of dust ($w=0$), the coordinate of the critical line (see Table \ref{table:1a}) is $\Omega=-2/3-\bar{\alpha}\Omega_0$
    and it tells us that the energy density $\Omega$ can be positive only when $\bar{\alpha}\Omega_0<-2/3$: as we always have $\Omega_0>0$, this requires $\bar{\alpha}<0$.  As for the stability, without imposing conditions on the curvature, we can state that in the case of dust the point is a center if $-1/3<\Omega_0 \bar{\alpha}<0$ and an unstable saddle if $\Omega_0\bar{\alpha}<-1/3$. Hence in the case of dust, due to the condition for positive energy density stated above, the line is an unstable saddle. Therefore, the allowed observational range $\bar{\alpha}\Omega_0=[-0.0223,0.0033]$ falls inside the intervals for which the critical line is a center.

    \item[$B_-$:] a particular case of the critical line above. It corresponds to a GR solution in which the energy density of the fluid is negative for any $w>-1$ and positive otherwise.
    
    \item[$C_-^\pm$:] negative curvature-dominated Milne solutions, i.e., with $a(t)\propto t$.  For $w<-1/3$ the expanding solution is a source, while the contracting one is a sink; they are both saddle points otherwise.
    
    \item[$A_+$:] line of points representing positive curvature-dominated, static solutions.  They are sourced by the modifications introduced by $\bar{\alpha}$.

Specializing to the case of dust, the coordinate of the critical line together with the condition of positive energy density gives $\bar{\alpha}\Omega_0<2/3$.  In this case we can have both conditions: $\bar{\alpha}$ either positive or negative.  In particular, imposing as well the positivity of $\Omega_0$ together with the theoretical constraint $-1/5<\bar{\alpha}<1$, we have that
\begin{align}
 -1/5<\bar{\alpha}<1\quad &\text{if}\quad \Omega_0\leq 2/3,\\
 -1/5<\bar{\alpha}<2/3\Omega_0\quad &\text{if}\quad \Omega_0>2/3.
\end{align}
As regards the stability, when we fix $w=0$, if $\bar{\alpha}<0$ the line behaves as a saddle for $-2/3<\bar{\alpha}\Omega_0<0$ and as a center $\bar{\alpha}\Omega_0 <-2/3$, while if $\bar{\alpha}>0$ it is a saddle for $0<\bar{\alpha}\Omega_0<1/3$ and a center for $\bar{\alpha}\Omega_0>1/3$.  The combined observational constraints give $\bar{\alpha}\Omega_0=-0.0095\pm 0.0128$, which can be satisfied for both positive or negative $\bar{\alpha}$. Therefore, the allowed observational range $\bar{\alpha}\Omega_0=[-0.0223,0.0033]$ falls inside the intervals for which the critical line is a saddle.
    
    \item[$B_+$:] a particular case of the critical line above with $\Omega_0=0$ or  $\bar{\alpha}=0$, describing positive curvature-dominated, static solutions.  They are saddle points for $w>-1/3$, and stable centers otherwise.
\end{itemize}

We see that the constant inertial mass density that comes from EMLG contributions gives rise to new interesting critical points, $B_0^\pm$ and $A_\pm$. As a consequence, it is also possible to have new trajectories that connect the new critical points with the GR ones.

\section{Discussion on $B_0^\pm$}
\label{sec:discussion}


In EMLG, from Eq.~\eqref{noncons}, we see that $\dot{\rho}=0$ can be achieved for $\Omega=\Omega_{\rm emlg}=\frac{2\bar{\alpha}}{\gamma}\Omega_0\frac{1+3w}{1+3w^2}$. Therefore there is no theoretical requirement for the source to have $w=-1$ in order to have a de Sitter behaviour. An example is represented by the class of solutions $B_0^+$, which describes a de Sitter future attractor due to EMLG modification, as an alternative to the attractor solution $A_0^{+}$ of $\Lambda$CDM, see Figures \ref{fig:attpos}/\ref{fig:attneg} for positive/negative $\bar{\alpha}$ values. Barrow and Board have been pointed out its existence in \cite{Board:2017ign}: for general EMSG theories, they expected to find a de Sitter solution to the modified Friedmann equation, the same as in GR except with altered constants. Here we have shown that such solution exists in EMLG,  as they predicted, and this is an attractor solution for both $w<-1$ and $w>-1/3$, just like a $\Lambda$-dominated solution for $w=-1$ in GR is.

\begin{figure}[t!]
\captionsetup{justification=raggedright,singlelinecheck=false}
\includegraphics[width=9.2cm]{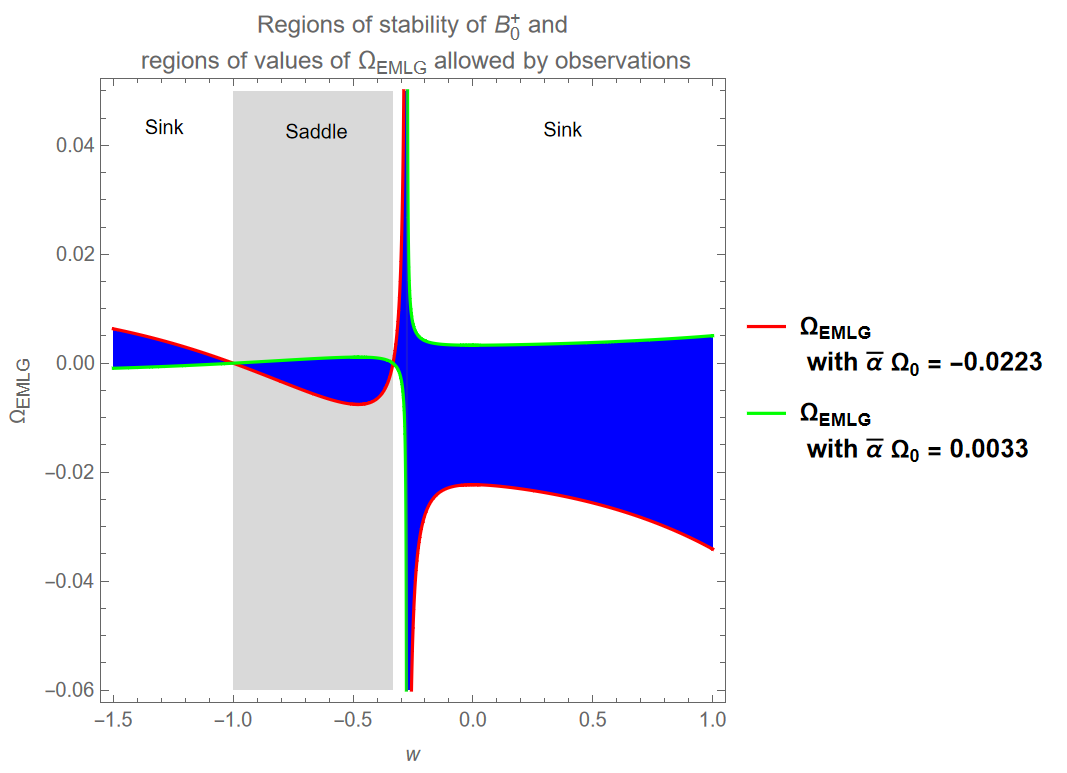}
    \caption{Stability of $B_0^\pm$ (grey (saddle) and white (sink) areas) via $\Omega_{i,\rm emlg}$ versus $w_{\rm i}$ plot with $\bar{\alpha}\Omega_0$ in the observationally allowed region.}
    \label{fig:imd1}
\end{figure}

In Figure \ref{fig:imd1}, we depict the EoS dependence of the critical point $B_0^+$'s stability with $\frac{\Omega_{\rm emlg}}{\bar{\alpha}\Omega_0}=\frac{2}{\gamma}\frac{1+3w}{1+3w^2}$. For phantom fluids $w<-1$, if $\bar{\alpha}<0$, the inertial mass density is positive ($\varrho>0$) and vice versa, and this solution is an attractor.  Another attractor solution branch is in the range $w>-1/3$: whereas usually any cosmological fluid with $w>-1/3$ causes deceleration of the universe, in EMLG a constant inertial mass density is generated which contributes to the acceleration. Such branch includes dust, and the sign of $\bar{\alpha}$ determines the sign of inertial mass density, i.e., for $\bar{\alpha}>0$ then $\varrho>0$, and for $\bar{\alpha}<0$ then $\varrho<0$. For quintessential sources $-1<w<-1/3$ (that also in GR cause acceleration) and $\bar{\alpha}<0$ with $\varrho<0$, the point is a saddle solution.

As can be seen from Figure \ref{fig:imd1}, for the special choice of $w=0$, i.e., dust, we thereby show that the analytical solution given in \cite{Akarsu:2019ygx} is a future attractor in the allowed observational range $\bar{\alpha}\Omega_0=[-0.0223,0.0033]$ (blue band).

Due to the additional inertial mass density of the pressureless matter, the evolution of the energy density gets modified as
\begin{equation}
\begin{aligned}
\label{soln}
\left[\frac{\rho_{\rm m}}{\rho_{\rm m0}}\right]_{\rm emlg}=\frac{1}{2}\left(\beta\pm\sqrt{\beta^2-4\bar{\alpha}\beta}\right)-\bar{\alpha},
\end{aligned}
\end{equation}
where $\beta\equiv(1+\bar{\alpha})^2 a^{-3}$ provided that $-1<\bar{\alpha}\leq1$. We remark that the case with negative sign in Eq.~\eqref{soln} does not have the GR limit as $\bar{\alpha}\rightarrow 0$ (in that case $\rho_{\rm m,emlg}\rightarrow 0$), but for the positive sign case, as $\bar{\alpha}\rightarrow0$, one recovers the standard $\Lambda$CDM model along with GR evolving with
\begin{equation}
\begin{aligned}
\left[\frac{\rho_{\rm m}}{\rho_{\rm m,0}}\right]_{\rm emlg}=\beta_{\rm GR}\equiv a^{-3},
\end{aligned}
\end{equation}
Differently than in GR, in EMLG theory, if the universe were to expand forever ($a \rightarrow \infty$ or $\beta \rightarrow 0$),  the energy density of pressureless matter would never reach zero, settling instead on a nonzero minimum constant, namely its inertial mass density with a negative sign, $\rho_{\rm m,emlg}\rightarrow -\bar{\alpha}\rho_{\rm m0,emlg}$ (or $\Omega_{\rm m,emlg}\rightarrow -\bar{\alpha}\Omega_{\rm m0,emlg}$).

The conservation equation \eqref{noncons} for dust takes the following form:
\begin{equation}
\label{noncons5}
\dot{\rho}_{\rm m}+3H\rho_{\rm m}\left[\frac{\rho_{\rm m}+\bar{\alpha}\rho_{\rm m0}}{\rho_{\rm m}-\bar{\alpha}\rho_{\rm m0}} \right]=0,
\end{equation}
The multiplier of $3H\rho_{\rm m}$ is null for $\Omega_{\rm m}\rightarrow -\bar{\alpha}\Omega_{\rm m0}$, which is the coordinate of the critical point $B_0^+$.
In this case we obtain $\dot{\rho}_{\rm m}\rightarrow0$ for $\rho_{\rm m}=-\bar{\alpha}\rho_{\rm m0}$, ($\Omega_{\rm m}= -\bar{\alpha}\Omega_{\rm m0}$).  Hence, the value of $\Lambda$ is altered by the constant inertial mass density of dust.  As $\Omega_{\rm m0}$ is always positive, the constant inertial mass density of dust $\bar{\alpha}\rho_{\rm m0}$ is positive (negative) when $\bar{\alpha}$, the coupling of the logarithmic contribution, is positive (negative). This enlightens us about the possibility of having negative values of inertial mass density for dust due to its modified conservation of energy, which is promising in view of alleviating in Hubble tension, as dust screens the $\Lambda$ at late times and concurrently contributes to $\Lambda$ altering the value of the cosmological constant in the future of the universe.

Here, accordingly, the Friedmann equation  \eqref{eq:rhoprime} reads:
\begin{equation}
\begin{aligned}  
\frac{H^2}{H_{0}^2}=\Omega_{\Lambda0}&+\Omega_{\rm m0}\bigg\{\frac{1}{2}\left(\beta+\sqrt{\beta^2-4\bar{\alpha}\beta}\right)\\
&-\bar{\alpha}\ln{\left[\frac{1}{2}\left(\beta+\sqrt{\beta^2-4\bar{\alpha}\beta}\right)-\bar{\alpha}\right]}\bigg\},
\label{Hpar}
\end{aligned}
\end{equation}
where $\Omega_{\Lambda0}=1-(1+\bar{\alpha}) \Omega_{\rm m0}$.\footnote{From \eqref{Hpar}, we can see that the EMLG model has no simple-graduated dark energy limit, as one recovers the standard $\Lambda$CDM model for $\bar{\alpha}=0$ (or $\beta=a^{-3}$). This is indeed consistent with our predictions as $\alpha=0$ in \eqref{action} with \eqref{eq:fchoice} would give Einstein-Hilbert action with a cosmological constant and matter action to describe dust. In the early Universe ($a\rightarrow 0$ or $\beta\rightarrow\infty$), EMLG converges to the simple-gDE model having Friedmann equation $H^2/H_{0}^2=\Omega_{\Lambda0}-\Omega_{\rm m0}6\bar{\alpha}\ln{(1+\bar{\alpha})\ln{(1+z)}}+\Omega_{\rm m0}(1+\bar{\alpha})^2(1+z)^{3}$, where the value of density parameter of dust today is rescaled with $\bar{\alpha}$; this seems to lead to a pseudo non-minimal interaction between the simple-gDE and dust.} 

The Hubble parameter reaches in the future a finite constant as in the $\Lambda$CDM model, but in this case not equal to $H_0\sqrt{\Omega_{\Lambda0}}$, instead equal to the altered constant due to the contribution of the constant inertial mass density of EMLG $H\rightarrow\,\,H_0\sqrt{\Omega_{\Lambda0}-\Omega_{\rm m0}\bar{\alpha}\ln{(-\bar{\alpha})}}$. The time derivative of $H$ still becomes null, i.e., $\dot{H}\rightarrow\,\,0$, leading to a de Sitter Universe, but with a modified cosmological constant. This result gives rise to the question whether $\Lambda$ should be included or not in this model, as $\Lambda$ is compulsory in this solution. Without $\Lambda$, the constraint on $\bar{\alpha}$ does contradict the allowed range for this solution,  $-1<\bar{\alpha}\leq1$ (see Section IIID of \cite{Akarsu:2019ygx} for the detailed explanation). 
For an eternally expanding universe, which is required to reach de Sitter universe in the future, $\bar{\alpha}$ must be negative, viz., the inertial mass density of dust must be negative, $\varrho_{\rm m}<0$. Now we hold a discussion on signature of $\alpha$ from a different point of view.

Future of universe depends on the sign of $\alpha$ and from \eqref{noncons5}, we see that the signature of EMLG contribution to energy conservation, viz., the sign of $\bar{\alpha}$, can be limited by adopting the irreversible thermodynamics of open systems in the presence of matter creation/annihilation. Extending the concept of adiabatic transformation (the heat transfer is not allowed, ${\rm d}Q=0$) from closed to open systems (which has been first done in \cite{Prigogine:1989zz}),
consider a system containing a non-constant number $N$ of particles in a volume element $V=a^3$: the thermodynamic energy conservation law is given by
\begin{align}
\label{eq:thermo}
{\rm d}(\rho V)+p{\rm d}V-\frac{\varrho}{n}{\rm d}(nV)=0,    
\end{align}
where $n$ is the particle number per unit volume and $\varrho=\rho+p$ is entalpy $h$ per unit volume, viz., the inertial mass density. While entropy change ${\rm d}S$ vanishes for an adiabatic transformation, in open systems 
\begin{align}
\label{eq:thermoS}
T{\rm d}S=\left(\frac{\varrho}{n}-\mu\right) {\rm d}(nV)=\frac{Ts}{n}{\rm d}(nV),    
\end{align}
where $\mu n=\varrho-T s$ is the chemical potential and the second law of thermodynamics requires that the total entropy of a system either increases or remains constant in any spontaneous process:
\begin{align}
{\rm d}(nV)\ge0.
\end{align}
 In \cite{Harko:2014pqa}, this is related to a process of creation of matter through the universe evolution in geometry-matter coupling theories.
Eq.~\eqref{eq:thermo} along with Eq.~\eqref{eq:thermoS} leads to
\begin{align}
\label{eq:def}
\dot{\rho}+3H\varrho=\varrho \left[\frac{\dot{n}}{n}+3H\right]=\varrho \frac{\dot{S}}{S}, 
\end{align}
Hence there is a possible contribution to entropy production due to non-conservation of energy density of dust with $\varrho_{\rm m}=\rho_{\rm m}$. Eqs.\eqref{noncons5} and \eqref{eq:def} give \footnote{If we assume that there is an apparent non-minimal interaction between simple gDE and dust, the conservation equation can be in such a way that $\nabla^{\mu}T_{\mu\nu}=Qu_{\nu}$ and $\nabla^{\mu}T_{\mu\nu}^{\rm emlg}=-Qu_{\nu}$. From \eqref{eq:def}  and \eqref{noncons5a} for dust, the energy-momentum transfer function $Q$ between the two components is given by $Q=\varrho_{\rm m}\frac{\dot{S}}{S}=\frac{12\alpha}{2\alpha-\rho_{\rm m}}H\rho_{\rm m}$ where $\rho_{\rm m}$ and $H$ are given by eqs.\eqref{soln} and \eqref{Hpar} respectively. The type of the interaction is generally of the form $Q\propto H\rho_{x}$ where $x$ denotes the energy density of dark matter or dark energy, whereas in the apparent non-minimal interaction interpretation of EMLG has a similar but more complicated form.}
\begin{equation}
\label{noncons5a}
\frac{\dot{S}}{S}=-3H\frac{4\alpha}{\rho_{\rm m}-2\alpha}\ge 0,
\end{equation}
where the expansion of universe viz., $H>0$ is achieved only for
\begin{align}
\alpha=2\bar{\alpha}\le0, 
\label{wemsg}
\end{align} 
giving negative inertial mass density, $\varrho_{\rm m}=2\alpha<0$. On top of that, the screening mechanism is only allowed for $\alpha<0$, consistent with the second law of thermodynamics. 

Figure \ref{fig:d} represents the parameter space for the lower bound on $\bar{\alpha}=-0.075$, in which the $\Omega_{H}=1$ slice of the three dimensional plot corresponds to the spatially flat Universe: the green dashed line is the critical line $B_0^+$ and represents the exponentially expanding solution, which is a sink; the solid green line is a trajectory that emerges from the dust-dominated source $C_0^+$ and ends in the exponentially expanding attractor $B_0^+$. Upgrading the null inertial mass density of the usual vacuum energy to an arbitrary constant in standard GR, the source satisfying $\varrho=\rm{const}$ has recently been of interest to many as it can resemble $\Lambda$ today, while leading to a future singularity dubbed the little sibling of the big rip (LSBR) for $\varrho=\rm{const}<0$ or a finite future bounce for $\varrho=\rm{const}>0$ \cite{Bouhmadi-Lopez:2014cca,Albarran:2016mdu,Bouali:2019whr}. On the other hand, in EMLG, the modified evolution of the dust leads to the fact that $\dot{H}$ still remains zero. While negative constant inertial mass density of dust does not lead to LSBR singularity as in the standard GR, a positive inertial mass density of dust ($\bar{\alpha}>0$) would lead that the scale factor infinity and it will have a bounce at the values for $a_*=\left[\frac{(1+\bar{\alpha})^2}{4\bar{\alpha}}\right]^\frac{1}{3}$ with $\rho_{\rm m*,emlg}=\bar{\alpha}\rho_{\rm m,0}$, namely, $\rho_{\rm m*,emlg}$ stops at a finite $a$, i.e.,  $a_*$, when $\beta=4\bar{\alpha}$. For a bounce to occur ($H_*=0$) $\bar{\alpha}=1.8$ is required but positive $\bar{\alpha}$ values are prohibited by the second law of thermodynamics, therefore the bouncing scenario in EMLG is not allowed. So we can see that the source leading the LSBR singularity/bouncing due to negative/positive inertial mass density in standard GR does not lead any singularity or bouncing in EMLG, and leads instead to a de Sitter future with a positive/negative constant deviation from $\Lambda$. In this gravity extension, matter (dust) has been modified while the geometry stays exactly the same as in GR: the modifications on the matter surprisingly disappear and the model seems the same as GR. This suggests that the different sources such as dark energy, Chaplygin gas, quintessence or phantom scalar fields should not be considered in GR, instead the coupling of matter to geometry should be modified with matter-based, alas logarithmic, modifications.

\section{Discussion on the critical points $A_\pm$ via $B_\pm$}
 A complete classification of all the Friedmann\textendash{}Lema\^\i{}tre\textendash{}Robertson\textendash{}Walker solutions with equation of state $w$ according to their conformal structure, singularities and trapping horizons has been given in \cite{Harada:2018ikn}. For a healthy discussion on the critical points $A_{\pm}$ of EMLG, we are first going to discuss $B_{\pm}$, the critical points of the GR system, for the cases $k>0$ and $k<0$ separately: 
\subsection{Positive spatial curvature $k>0$}
The critical point $B_+$ has coordinate $\Omega\equiv\frac{\rho}{3D^2}=\frac{2}{3(1+w)}$, giving the Friedmann equation 
\begin{equation}
\begin{aligned}
3D^2=\frac{3(1+w)}{2}\rho,
\end{aligned}
\end{equation}
the corresponding energy density evolves as 
\begin{equation}
\begin{aligned}
\label{eq:spati}
\rho\propto  a^{-3(1+w_{\rm eff})} \quad \textnormal{where}\quad 
w_{\rm eff}=-\frac{1}{3}+\frac{1+3w}{3+3w},
\end{aligned}
\end{equation}
where the source with $w=-1/3$ and $w_{\rm eff}=-1/3$ preserves the evolution of energy density $\rho\propto  a^{-2}$, and corresponding Friedmann equation reads
\begin{equation}
\begin{aligned}
\label{eq:friedspati}
H^2=\left(\frac{\rho_0}{3}-k\right)\left(\frac{a_0}{a}\right)^{2}=-\dot{H},
\end{aligned}
\end{equation}
which has two solutions. Here the considered solution is achieved if $\rho_0=3k$, then $a=a_0$ and $\eta=t/a_0$ is the conformal time where $a_0$ is a constant of integration. For closed space, the spacetime is then identical to the Einstein static universe ($H=\dot{H}=0$), with no singularity, and the domain of $\eta$ is $-\infty<\eta<\infty$. 

As explicitly seen from Eq.~\eqref{eq:spati}, there is an equivalence between the energy density of cosmic strings ($w=-1/3$) and the corresponding energy density of the spatial curvature in standard GR. In EMLG, from Eq.~\eqref{noncons}, we see that for $\bar{\alpha}\neq0$, this equivalence is broken and the energy density of cosmic string ($w=-1/3$) evolves as 
\begin{equation}
\begin{aligned}
\label{solncurv}
\rho_{\rm str,emlg}=\frac{\rho_{\rm str0,emlg}}{a^2}\exp{\left({\rm LambertW}\left[-\frac{2\bar{\alpha}\rho_{\rm str0,emlg}^2a^2}{\ln{(3/4)}}\right]\right)},
\end{aligned}
\end{equation}
valid for $\bar{\alpha}<0$, leading a slightly different evolution compared to GR, which has $\rho_{\rm str}\propto a^{-2}$ ($\bar{\alpha}=0$). 
Therefore, when compared to $B_+$, we expect to have different dynamics belonging to $A_+$, on the other hand, even the equality in GR is broken due to EMLG contributions, interestingly, $A_+$ CP has  $\Omega=\frac{2}{3(1+w)}+\frac{2 \bar{\alpha}(1+3w)}{\gamma(1+3w^2)}\Omega_0$ coordinate and Eq.~\eqref{noncons} gives
\begin{equation}
\begin{aligned}
\dot{\rho}+2H\rho=0 \quad \textnormal{leading} \quad \rho\propto a^{-2},
\end{aligned}
\end{equation}
irrespective of the type of the source. The source contributes to field equations as a positive curvature and the Friedmann equation has the same form with Eq.~\eqref{eq:friedspati}. There arises a new static solution in EMLG, where the source is arbitrary although interestingly acting as spatial curvature.
Given the allowed observational range $\bar{\alpha}\Omega_0=[-0.0223,0]$ along with the negativity condition, we conclude that the critical line is a saddle.\footnote{We now speculate about the curved space dynamics, indeed the observational analysis of EMLG presented in \cite{Akarsu:2019ygx} should be done for spatially curved universes as well to obtain a healthy discussion. }

\subsection{Negative spatial curvature $k<0$ with $\rho>0$}

The critical point $B_-$ has coordinate $\Omega\equiv\frac{\rho}{3D^2}=-\frac{2}{3(1+w)}$,  giving the Friedmann equation
\begin{equation}
\begin{aligned}
3D^2=-\frac{3(1+w)}{2}\rho.
\end{aligned}
\end{equation} 
Here at first sight, energy density seems negative. The corresponding energy density evolves as 
\begin{equation}
\begin{aligned}
\label{eq:spatiopen}
\rho\propto  a^{-3(1+w_{\rm eff})} \quad \textnormal{where}\quad 
w_{\rm eff}=-\frac{5}{3}+\frac{5+3w}{3+3w}.
\end{aligned}
\end{equation}
Here, as can be seen from 
\begin{equation}
\begin{aligned}
H^2=&\frac{\rho_0}{3}\left(\frac{a_0}{a}\right)^{3(1+w_{\rm eff})}-k\left(\frac{a_0}{a}\right)^{2}, 
\end{aligned}
\end{equation}
a negative energy density is only possible for $k<0$, which describes a bouncing universe with future and past null infinities for $w>-1/3$, a universe beginning with a big-bang singularity and ending with a big-crunch singularity for $-1<w<-1/3$, and a universe emerging from a regular null hypersurface and then ending into another regular hypersurface for $w<-1$ (see \cite{Bouhmadi-Lopez:2019zvz} for details). But in $B_-$, the energy density is positive in Friedmann equation,  in terms of dimensionless variables used as in the Section \ref{sec:dyn},   
Eq.~\eqref{eq:rayneg} reads
for $\Omega=-\frac{2}{3(1+w)}$
  \begin{align}
    1 =& 2\, \Omega_H^2\,+\frac{2}{3(1+w)}, \label{eq:friedneg2}\\
    \frac{\dot{H}}{D^2}=&\,\Omega_H^2\equiv\frac{H^2}{D^2},
    \label{eq:rayneg2}
  \end{align}
corresponding deceleration parameter and $w_{\rm eff}$ for critical point $A_-$ are
\begin{align}
\label{eq:qdef}
 q = -1-\frac{\dot{H}}{H^2}=-2 \quad,\quad  w_{\rm eff} =-1-\frac{2}{3}\frac{\dot{H}}{H^2} =-\frac{5}{3}
\end{align}
which requires  $w=-5/3$, $\Omega_{H}=0$ (where $\dot{H}=H^2$ so their ratio equals to unity but $D\rightarrow\infty$ leading $\Omega_{H}=0$). For this case, energy density evolves phantom-like, $\rho\propto  a^{2}$, and Friedmann and Raychaudhuri equations are respectively as follows:
\begin{equation}
\begin{aligned}
\label{eq:friedspati2}
H^2=&\frac{\rho_0}{3}\left(\frac{a}{a_0}\right)^{2}-k\left(\frac{a_0}{a}\right)^{2}, \\
\dot{H}=&\frac{\rho_0}{3}\left(\frac{a}{a_0}\right)^{2}+k\left(\frac{a_0}{a}\right)^{2}, 
\end{aligned}
\end{equation}
leading to
\begin{equation}
\begin{aligned}
H^2+\dot{H}\equiv\frac{\ddot{a}}{a}=&\frac{2\rho_0}{3}\left(\frac{a}{a_0}\right)^{2}.
\end{aligned}
\end{equation}
Accordingly, the size of the observable universe becomes infinite at a finite time from present epoch, namely at $t_{\textup{rip}}$. This is, for a general $w$,
	\begin{eqnarray}
		t_{\textup{rip}}\coloneqq t_\star-\frac{2}{\sqrt{3\rho_0}(1+w)}\left(\frac{a_\star}{a_0}\right)^{\frac{3(1+w)}{2}}.
	\end{eqnarray}
Furthermore, given that
	\begin{align}
		H=&-\frac{2}{3(1+w)\left(t_{\textup{rip}}-t\right)}=\frac{\Omega}{\left(t_{\textup{rip}}-t\right)}
		,\label{eq:H alpha 1}\\
		\dot{H}=&-\frac{2}{3(1+w)\left(t_{\textup{rip}}-t\right)^2}=\frac{\Omega}{\left(t_{\textup{rip}}-t\right)}
		\label{eq:dotH alpha 1},
	\end{align}
the Hubble rate and its cosmic time derivative also diverge at $t=t_{\textup{rip}}$. For $w=-5/3$ we have $\Omega=1$. Therefore, the universe evolves towards a classical Big Rip singularity. This corresponds to a type I singularity according to the notation in Ref. \cite{EOSalpha3}. 


The critical point $A_-$ has $\Omega=-\frac{2}{3(1+w)}+\frac{2 \bar{\alpha}(1+3w)}{\gamma(1+3w^2)}\Omega_0$ critical coordinate and Eq.~\eqref{noncons} gives
\begin{equation}
\begin{aligned}
\dot{\rho}-2H\rho=0 \quad \textnormal{leading to} \quad \rho\propto a^{2},
\end{aligned}
\end{equation}
irrespective of $w$, i.e., the type of the source, and $\bar\alpha$. It is worth noting that the source in this case, contributes to the field equations like a phantom field independent of the value of EoS parameter, and there arises a new solution in EMLG. And, in the asymptotic future, $\Omega$ still depends on not only $w$ but also $\bar{\alpha}$. Hence the stability of this point depends on $\bar{\alpha}\Omega_0$: if we consider dust and the observational allowed range along with the negativity condition on $\bar{\alpha}$, $\bar{\alpha}\Omega_0=[-0.0223,0]$, we conclude that this point is a center.

As shown in Figure \ref{fig:d}, when the lower observational bound on $\bar{\alpha}=-0.075$ is considered, in comparison to GR (where $\bar{\alpha}=0$, the upper bound on $\bar{\alpha}$), we see new trajectories corresponding to recollapsing models (blue in the plot). This means that in EMLG, the transition from $C_-^\pm$ critical points is allowed due to $\bar{\alpha}\neq0$, while it is forbidden in GR ($\bar{\alpha}=0$) shown in Figure \ref{fig:c}.\\

\section{Conclusions}
We have carried out the dynamical system analysis of a cosmological model in the energy-momentum squared gravity (EMSG) described by the functional $f(T_{\mu\nu}T^{\mu\nu})=\alpha\ln({\lambda}T_{\mu\nu} T^{\mu\nu})$, which is known as energy-momentum log gravity (EMLG) \cite{Akarsu:2019ygx}. In this model, the new terms in the right-hand side of the Einstein field equations yield a constant inertial mass density and provide a dynamical dark energy  with a density passing below zero at large redshifts, accommodating a mechanism for screening $\Lambda$ in the past, suggested for alleviating some cosmological tensions such as the $H_0$ tension \cite{Akarsu:2019ygx} (see also \cite{Delubac:2014aqe,Sahni:2014ooa,Aubourg:2014yra,Capozziello:2018jya,Wang:2018fng,Poulin:2018zxs,Dutta:2018vmq,Banihashemi:2018oxo,Visinelli:2019qqu,Akarsu:2019hmw,DiValentino:2020naf,Akarsu:2021fol,Escamilla:2021uoj,DiValentino:2020zio,DiValentino:2021izs}. 

We have shown that the analytical cosmological solution of EMLG presented in \cite{Akarsu:2019ygx} is a future attractor; see the critical point $B_0^+$ for dust $(w=0)$. In particular, the model gives rise to an entire class of new stable late-time solutions with  $H\rightarrow\sqrt{(\Lambda+2\alpha)/3}$ as $a\rightarrow\infty$, where the new term is due to the constant effective inertial mass density that arises from EMLG contribution of dust (which should be negative, since $\alpha<0$ is required by the second law of thermodynamics), whereas $H\rightarrow\sqrt{\Lambda/3}$ as $a\rightarrow\infty$ in the standard $\Lambda$CDM model. In EMLG, we see that $\dot{\rho}=0$ can be achieved for $\Omega=\Omega_{\rm emlg}=\frac{2\bar{\alpha}}{\gamma}\Omega_0\frac{1+3w}{1+3w^2}$ for any type of source, that is irrespective of the EoS parameter of the source $w$: this class of solutions $B_0^+$ which describes a de Sitter future attractor due to EMLG modification has been obtained as an extension of the attractor solution $A_0^{+}$ of $\Lambda$CDM.

We have also shown that the presence of an EMLG modification, on top of GR, with an appropriate EoS parameter, can reproduce the effects of spatial curvature (see critical point $B_+$) and phantom fluid (see critical point $B_-$). Moreover, in contrast to GR, as it is clear from Fig.~\ref{fig:d}, the EMLG extension of GR allows for recollapsing universe dynamics even for the spatially open RW spacetime (the negative curvature case).

\begin{acknowledgments}
The authors thank to \" Ozg\" ur Akarsu for valuable discussions. N.K. thanks Do\u gu\c s University for the financial support provided by the Scientific Research (BAP) project number 2021-22-D1-B01. N.K. acknowledges the COST Action CA21136 (CosmoVerse).
\end{acknowledgments}

\end{document}